# Novel Method for Mutational Disease Prediction using Bioinformatics Techniques and Backpropagation Algorithm

Ayad Ghany Ismaeel
Department of Information System Engineering, Erbil Technical College, Hawler Polytechnic University (previous FTE- Erbil), Iraq.
E-mail: dr_a_gh_i@yahoo.com alternative
dr.ayad.ghany.ismaeel@gmail.com

Anar Auda Ablahad
Department of Electrical and Computer, School of Engineering, Faculty of Engineering and Applied Science, University of Dohuk, Iraq.
E-mail: anar_alkasyonan@yahoo.com

*Abstract*— Cancer is one of the most feared diseases in the world it has increased disturbingly and breast cancer occurs in one out of eight women, the prediction of malignancies plays essential roles not only in revealing human genome, but also in discovering effective prevention and treatment of cancers. Generally cancer disease driven by somatic mutations in an individual's DNA sequence, or genome that accumulates during the person's lifetime. This paper is proposed a novel method can predict the disease by mutations despite The presence in gene sequence is not necessary it are malignant, so will be compare the patient's protein with the gene's protein of disease if there is difference between these two proteins then can say there is malignant mutations. This method will use bioinformatics techniques like FASTA, CLUSTALW, etc which shows whether malignant mutations or not, then training the back-propagation algorithm using all expected malignant mutations for a certain disease's genes (e.g. BRCA1 and BRCA2), and using it to test whether patient is holder the disease or not. Implementing this novel method as the first way to predict the disease based on mutations in the sequence of the gene that causes the disease shows two decisions are achieved successfully, the first diagnose whether the patient has cancer's mutations or not using bioinformatics techniques the second classifying these mutations are related to breast cancer (e.g. BRCA1and BRCA2) using back-propagation with mean square rate 0.0000001.

I. INTRODUCTION

Cancer is a leading cause of death worldwide; it has accounted for 7.4 million deaths (around 13% of all deaths) in 2004, with an estimated 12 million deaths in 2030 [1], breast cancer is the second leading cause of cancer deaths worldwide. The malignant tumor develops when cells in the breast tissue divide and grow without the normal controls on cell death and cell division. It is the most common form of cancer in females that is affecting approximately 10% of all women at some stage of their life. These factors include such attributes as age, genetic risk and family history, but the survival rate is high with early diagnosis, 97% of women survive for 5 years or more years [2]. Predicting malignancies plays essential roles not only in revealing human genome, but also in discovering effective prevention and treatment of cancers, curing from this malicious disease is mainly based upon early and accurate diagnosis computer aided diagnosis can be a good helper in this area.

Cancer is a disease driven by somatic mutations in an individual's DNA sequence, or genome that accumulates during the person's lifetime. These mutations arise during DNA replication, which occurs as cells grow and divide into two daughter cells. Mutations arise as errors in the DNA replication process and distinguish the DNA in the daughter cells from the parental cells. They take place on a continuum of scales ranging from single "character" substitutions (the nucleotides Adenine A, Thymine T, Cytosine C, and Guanine G of DNA) to structural variants that duplicate, delete, or rearrange larger genome segments [3].

So there is a big problem needs to be solve, through proposed a novel method to predict the disease by its mutations which discover on patient's gene sequence and its protein using two important categories in this field which are bioinformatics technique and back-propagation BP algorithm. The following demonstration will reveal why selecting these categories in this proposed new method.

Today's world, computers are as likely to be used by biologists as by any other highly trained professionals, however besides these general tasks, biologists also use computers to address problems that are very specific to biologists these specialized tasks, taken together, make up the field of bioinformatics. More specifically can define bioinformatics; its focus is to make predictions about biological systems and to analyze biological data in an effort to provide more insight into how living organisms function. For example, computational biology could be used to predict if two proteins interact or not, if prediction is accurate then computational biology could further be used to analyze biological data obtained from a wet lab experiment involving the proteins to understand how these proteins contribute to physiology of organism. Computational biology can be further broken down into molecular modeling and bioinformatics [4].

At the last decade, using of artificial intelligence AI become widely accepted in medical applications [5] an artificial neural network ANN is defined as an information processing system inspired by the structure of the human brain. ANN gathers its knowledge by detecting a common pattern and relationships in raw data, the learning from such relationships and adapting the results as required. It is composed of a large number of highly interconnected processing elements that are analogous to synapses, ANN used to predict and learn from a given sets of data [6].

## II. RELATED WORK

Muhammad S. and Zaihisma C. [2010] develop a system that can classify "Breast Cancer Disease" tumor using neural network with feed forward back-propagation algorithm to classify the tumor from a symptom that causes the breast cancer disease. The aim of research is to develop more cost-effective and easy–to-use systems for supporting clinicians. This study, proposed a back-propagation in breast cancer diagnosis problem. This method consists of two-stages. First stage, the input data is trained by using feed-forward rules, in the second stage neural network model is used to classify the breast cancer data. Experimental results show that the concise models extracted from the network achieve high accuracy rate (0.001) of on the training data set and on the test data set. Breast cancer tumor database used for this purpose is from the University of Wisconsin, machine learning repository [6].

Shweta Kharya [2012] discussed various data mining approaches that have been utilized for breast cancer diagnosis and prognosis. Breast cancer diagnosis is distinguishing of benign from malignant breast lumps and breast cancer prognosis predicts when breast cancer is to recur in patients that have had their cancers excised. This method summarizes various reviews and technical articles on breast cancer diagnosis and prognosis also will focus on current research being carried out using the data mining techniques to enhance the breast cancer diagnosis and prognosis, additional the decision tree which used found best predictor (93.62%) accuracy [2].

Steven L Salzberg and Mihaela Pertea [2010] developed a computational screen that tests an individual's genome for mutations in the BRCA genes because currently Myriad charges more than $3000 for its tests on the BRCA genes, while sequencing one's entire genome now costs less than $20,000, despite the fact that both are currently protected by patents, i.e. made the software freely available (at http://cbcb.umd.edu/software/BRCA-diagnostic) [7] .

Huseyin K. and Sule G. Oguducu [2010], proposed a new method to use in diagnostic process of genetic disorders to determine the mutations in DNA sequences. The contribution of method is used chromatograms without applying a base calling method in order to decrease the errors produced during the base calling step. Given reference and unknown chromatograms, this method searches for possible mutations in the unknown chromatogram against the reference one, the approach first extracts feature vectors of both chromatograms by applying a two dimensional transformation to every data frame sliding through the chromatograms. Test of method on a freely available dataset shows results a new method can successfully align two chromatograms and highlight the differences caused by mutations [8].

The weakness of all these methods and techniques they focus in diagnostic or prediction base on features of mutations in disease's genes not genome sequence comparing with patient's gene, with taking into consideration depending on the idea of "two sequences may have big differences in DNA sequence but have similar protein" [10].

The motivation reach to a novel method for prediction is the patient has disease base on genes mutations which caused the disease and not that only, but also will transition to checkup their proteins, i.e. check the mismatching, e.g. between the breast cancer genes (BRCA1 & BRCA2) and their proteins with the patient's gene sequence and it's protein to find malignancy mutations using bioinformatics technique at first, then when find these malignancy mutations will use trainer BP algorithm with all gene's mutations that causes the disease then classify the mutations in patient's gene whether related to BRCA1/BRCA2 which caused breast cancer or not.

## III. PROPOSED OF NOVEL METHOD FOR DISEASE PREDICTION

The main tasks of suggested novel method shown in Fig. 1, this method of disease prediction needs two approaches; first whether the patient has mutations causes the disease or not, the second discover these mutations are related to certain disease. Those two approaches are:

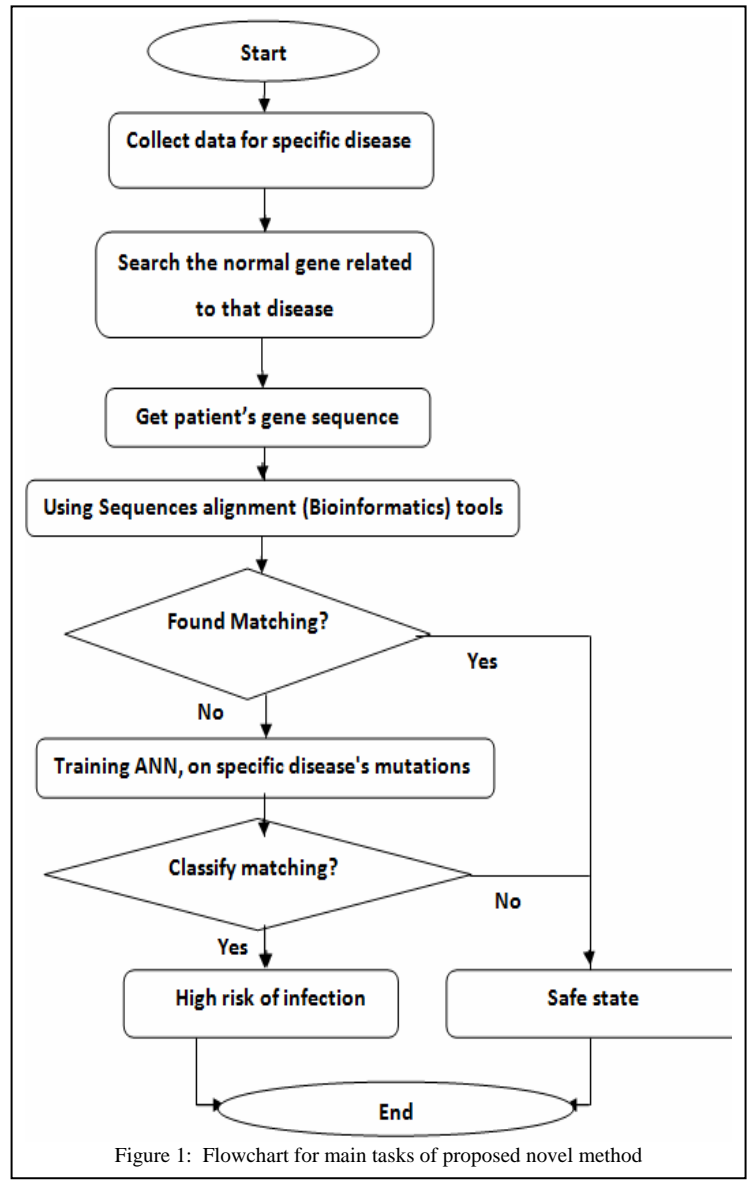

Figure 1: Flowchart for main tasks of proposed novel method

IRACST – Engineering Science and Technology: An International Journal (ESTIJ), ISSN: 2250-3498,
Vol.3, No.1, February 2013*A. Bioinformatics techniques*

Need to know how to access and get helpful bioinformatics tools for genome search, analysis, etc in following sequence:

1) FASTA: Compares a query string against a single text string when searching the whole database for matching for given query, that comparing done by using the FASTA algorithm to every string in the database. When looking for an alignment might expect to find a few segments in which there will be absolute identity between the two compared strings, FASTA used this property and focus on identical regions.

2) CLUSTALW: it is first technique for checkup whether the patient has malignant mutation for cancer or no, and depending on the idea of "Two proteins can have very different amino acid sequences and still be biologically similar (Homology)" [10], CLUSTALW help in detection the gene mutation which increasing probability of cancer according to number of genes that are related with cancer (such as brca1 , brca2 which cause the breast cancer), in CLUSTALW must be knowing the normal sequence of each gene (without annotation) with patient's gene sequence and check the matching between them, Fig.2 shows flowchart for the main tasks of the tools of sequences alignment (bioinformatics).

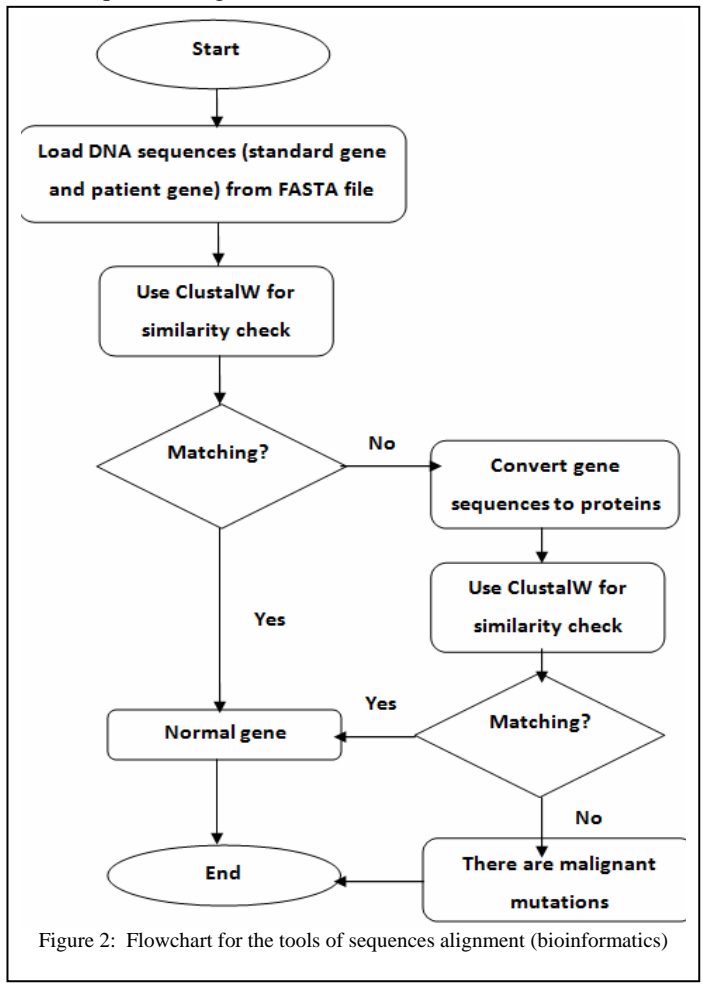

Figure 2: Flowchart for the tools of sequences alignment (bioinformatics)

*B. Back-propagation algorithm*

The results which obtained in previous approach (bioinformatics technique), shows there is malignant mutations or benign (normal) in the patient's genes, when there is malignant mutations must diagnosing those mutations related to a certain disease. So the classifying of mutations (e.g. which are related to breast cancer) needs another approach focus on training back-propagation, which most commonly used in medical research. BP networks where the signals travel in one direction from input neuron to an output neuron without returning to its source, neural network is a sorted topology which include node and weighted connections. Each layer inside neural network will be connected by their weight connector; BP network consists of at least three layers of units (input layer, at least one hidden layer and output layer) as shown in Fig. 3 [6].

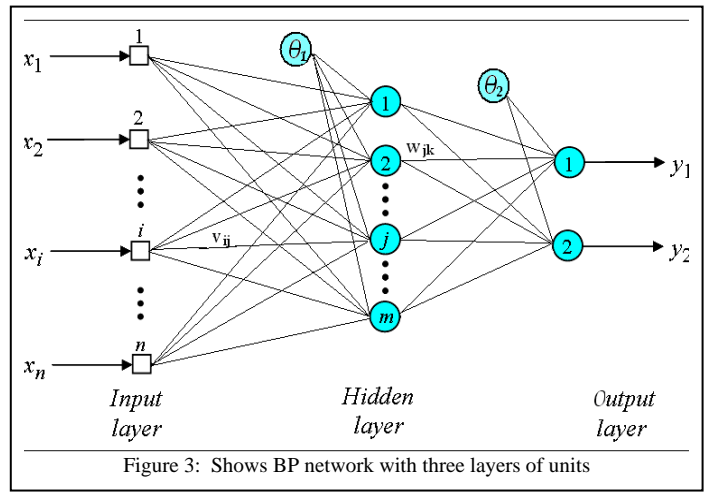

Figure 3: Shows BP network with three layers of units

That means the novel method will be used a feed forward back-propagation network as popular model in neural network, which hasn't feedback connections but the errors are back propagated during training. Training the feed forward BP algorithm used to create the best neural network model, this algorithm is divided into three main parts which are feed forward, error calculation and the last part is updating the weight, training algorithm presented below [6, 9]:

1) Each input unit ($X_i$, i = 1, 2, 3 …, n) receive input signal $x_i$, and sent the signal to all unit in their above layer (hidden units).

2) Each hidden layer unit ($Z_j$, j = 1, 2, 3 …, p) sum the input weight signal,

$$z\_in_j = v_{0j} + \sum_{i=1}^{n} x_i v_{ij}, \quad (1)$$

with using its activation function to get output signal value

$$z_j = f(z\_in_j), \quad (2)$$

and sent that signal to all units at its above layer (output units).

152

3) Each unit output ($Y_k$, k = 1, 2, 3 …, m) sum input weight signal,

$$y\_in_k = w_{0k} + \sum_{j=1}^{p} z_j w_{jk} \qquad (3)$$

with using its activation function to get output signal value,

$$y_k = f(y\_in_k). \qquad (4)$$

Feed forward process is the first part in back-propagation algorithm, which is used to send input signal into above layer and the next step is to do error calculation.

4) Each output unit ($y_k$, k = 1, 2, 3 …, m) accept one pair target pattern with input training pattern, count an error,

$$\delta_k = (t_k - y_k)f'(y\_in_k), \qquad (5)$$

count weight correction (used in update weight $w_{jk}$),

$$\Delta w_{jk} = \alpha \delta_k z_j, \qquad (6)$$

count bias correction (used in update bias value $w_{0,k}$)

$$\Delta w_{0k} = \alpha \delta_k, \qquad (7)$$

and sent $\delta_k$ to unit below it.

5) Each hidden unit ($Z_j$, j = 1, 2, 3 …, p) sum delta input (from its above layer units),

$$\delta\_in_j = \sum_{k=1}^{m} \delta_k w_{jk}, \qquad (8)$$

multiply with activation output to count an error,

$$\delta_j = \delta\_in_j f'(z\_in_j), \qquad (9)$$

count weight error (used for update $v_{ij}$);

$$\Delta v_{ij} = \alpha \delta_j x_i, \qquad (10)$$

and count its bias correction (used for update $v_{0j}$);

$$\Delta v_{0j} = \alpha \delta_j. \qquad (11)$$

## IV. EXPERIMENTAL RESULTS

Implementing the proposed novel method using breast cancer genes (BRCA1 & BRCA2) as second leading cancer cause deaths via two approaches as referring to in subsection 3:

### A. First approach

Need to use and get helpful bioinformatics tools for genome search, analysis, etc as below:

1) FASTA: This algorithm looking for Normal BRCA1 and BRCA2 genes in database [12], and expect to find a few segments in which there will be absolute identity that is done as follow:

Select NCBI web→ Type gene name→ Search across database→ Select type of organism→ Click FASTA, then→ finally will get what the search for it, as in Fig. 4. Addition the patient's gene must be ready for next step.

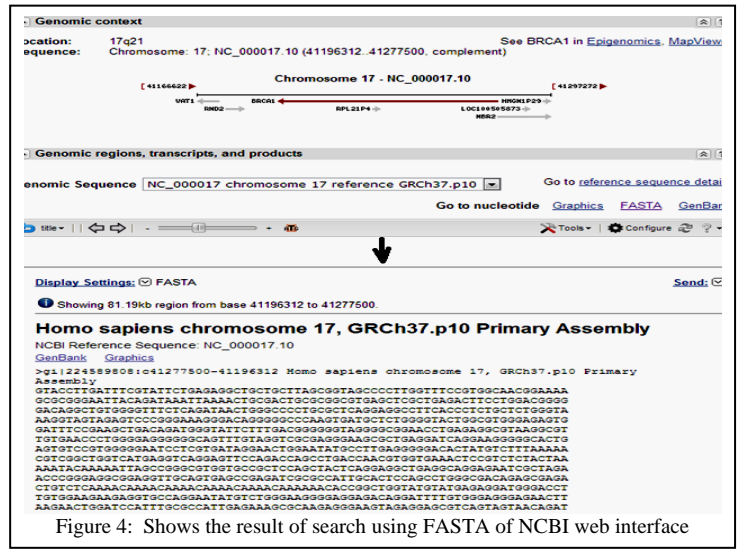

Figure 4: Shows the result of search using FASTA of NCBI web interface

2) CLUSTALW: This step will help in detection the affected gene mutations that increase probability of cancer. According to number of genes that are related to breast cancer such as BRCA1 and BRCA2 will implement CLUSTALW using BioEdit package which can accept Notepad file contain the standard (normal) gene sequence of BRCA1 or BRCA2 (gene without annotation or mutation that is used as the reference gene in the analysis), and patient's genes (patient's BRCA1 or BRCA2) to checkup whether mismatch among them or not. Fig. 5 shows example of found mismatch between two gene sequences alignment (sequence of BRCA2 and patient's sequence).

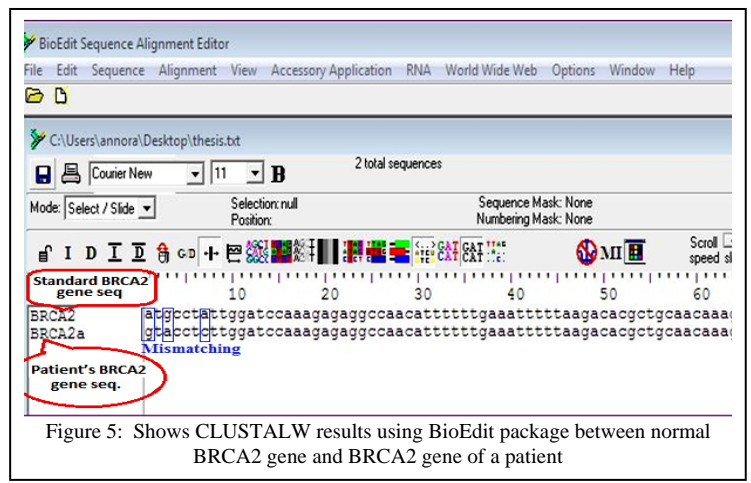

Figure 5: Shows CLUSTALW results using BioEdit package between normal BRCA2 gene and BRCA2 gene of a patient

Because there is mismatch as appears in Fig. 5, will transfer from BRCA2 gene sequence to its protein, then will checkup the mismatch between protein of normal BRCA2 and BRCA2's protein of patient if there is mismatching that means there are malignant mutations, Fig. 6 shows the mismatch.

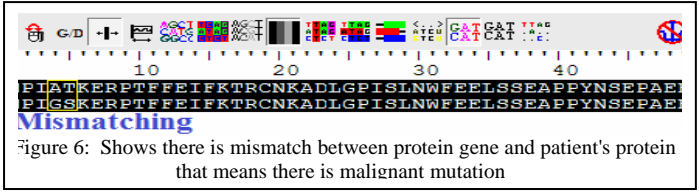

Figure 6: Shows there is mismatch between protein gene and patient's protein that means there is malignant mutation

### B. Second approach

Previous approach not enough for prediction because the malignant mutations which discovered are general, need to classify these mutations are related with gene of BRCA1 or BRCA2, so will use MATLAB R2010a on PC type Core i3 for neural network toolbox because it contains various functions can use for implementing feed forward neural networks. There are various back-propagation training algorithms with diverse capability based on the nature of problem the network is designed to solve. Generally, ANN can be trained for tasks such as function approximation (regression) or pattern recognition (discriminate analysis). At this novel method will focus on pattern recognition in which the network is trained with the nucleotide sequences of breast cancerous, i.e. depending on all malignant mutations in breast cancer which are 9 (5 mutations related with BRCA1 and 4 for BRCA2) which are compiled from the OMIM database of human genetic diseases and the literature survey [11], will training the feed forward back-propagation algorithm with mean square rate (net.trainParam.goal) = 0.0000001, this BP model has the following layers:

- Input layer with refer to mutations
- Two Hidden layers
- Output layer which notify normal gene or abnormal.

Fig. 7 shows the training result of feed foreword BP algorithm.

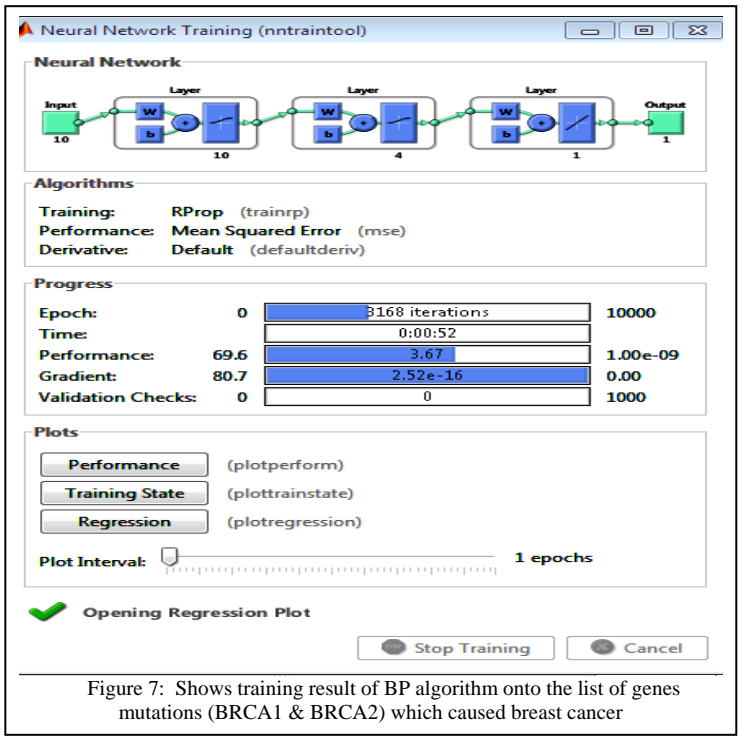

Figure 7: Shows training result of BP algorithm onto the list of genes mutations (BRCA1 & BRCA2) which caused breast cancer

Fig. 8 reveals plots of the important three elements of this training (Fig. 8; A shows performance, Fig. 8; B reveals training state and Fig 8, C shows regression).

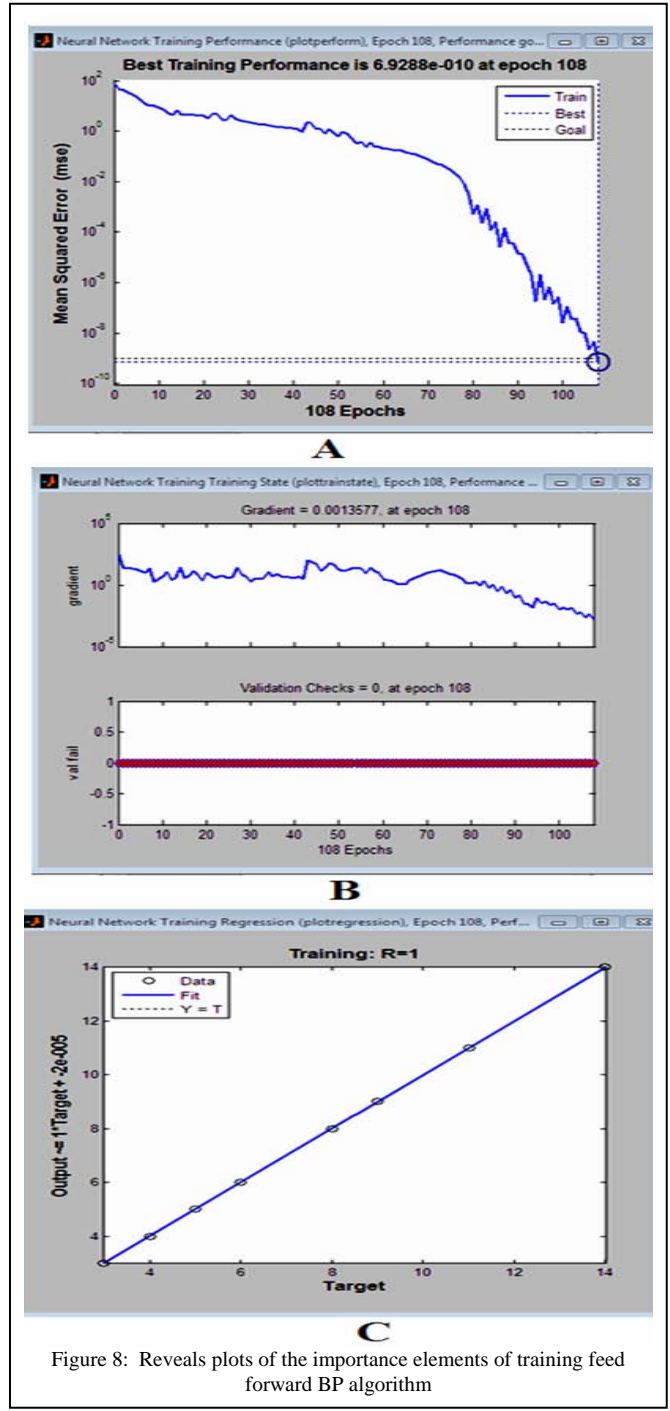

Figure 8: Reveals plots of the importance elements of training feed forward BP algorithm

Test the trainer BP with malignant mutations (9 mutations) of BRCA1 and BRCA2 that achieved using GUI designed for the users of novel method; Fig.9; A shows example, there isn't match between malignant mutations of BRCA1 (which are 5) and patient's gene (BRCA1), i.e. BP algorithm is classifying (predicted) the patient isn't holder of the disease, but Fig.9; B shows there is matching at 4th malignant mutation (The four candidate position) from the five mutations of BRAC1, i.e. the patient is holder of the disease (predicted the patient is abnormal).

*C. Discussion the results*

Comparing the results of implementing the novel method with other related techniques shown in Table 1:

TABLE I. REVEALS A COMPARISON OF PROPOSED NOVEL METHOD WITH OTHER TECHNIQUES/ METHODS

| Feature | Proposed of Novel Method | Muhammad S. and Zaihisma C. method [6] | Shweta Kharya method [2] |
|---|---|---|---|
| Using dataset sequences gene (DNA) and its protein. | Yes | No | No |
| Mean square rate or best predictor. | Mean square rate 0.0000001 | Mean square rate 0.001 | Best predictor 93.62% |
| Technique which is used for predicting a disease | Used bioinformatics techniques, additional to train feed forward BP to classify the malignant mutations of breast cancer at genes (which caused cancer) and their proteins. | Used feed forward BP to classify tumor from symptom that causes breast cancer. | Used data mining to enhance the breast cancer diagnosis & prognosis. |
| Universal method or technique. | Yes, if training BP with mutations of other diseases. | No | No |
| The results of novel method can use to support. | Researchers in bioinformatics, molecular biology, Biomedical, etc. | Researchers are limited, because it not works with gene & protein sequences. | Researchers are limited, because it not works with gene & protein sequences. |
| Cost-effective for analysis | Lower cost comparing to other methods of analysis, where the cost of the least one, about $3000 [7]. | Does not have this capability | Does not have this capability |

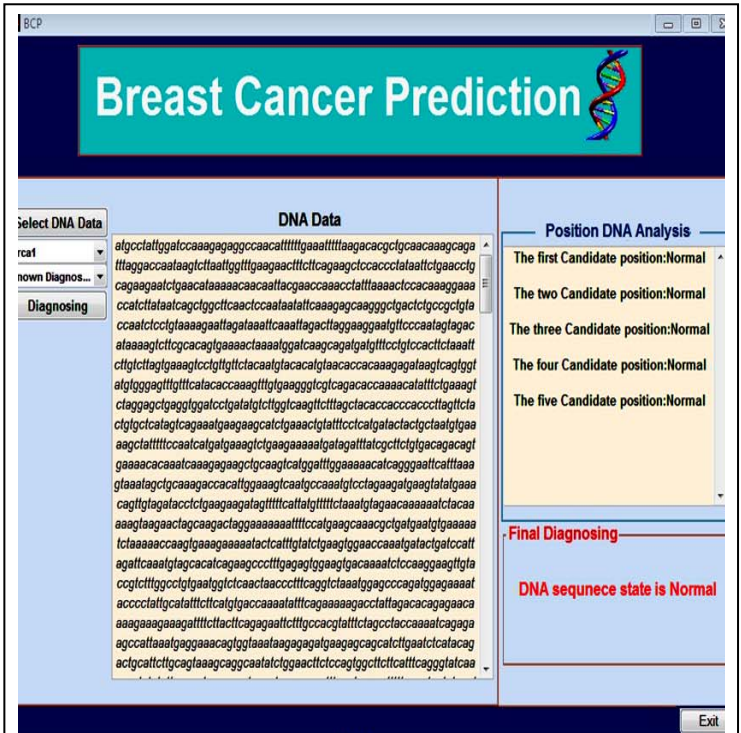

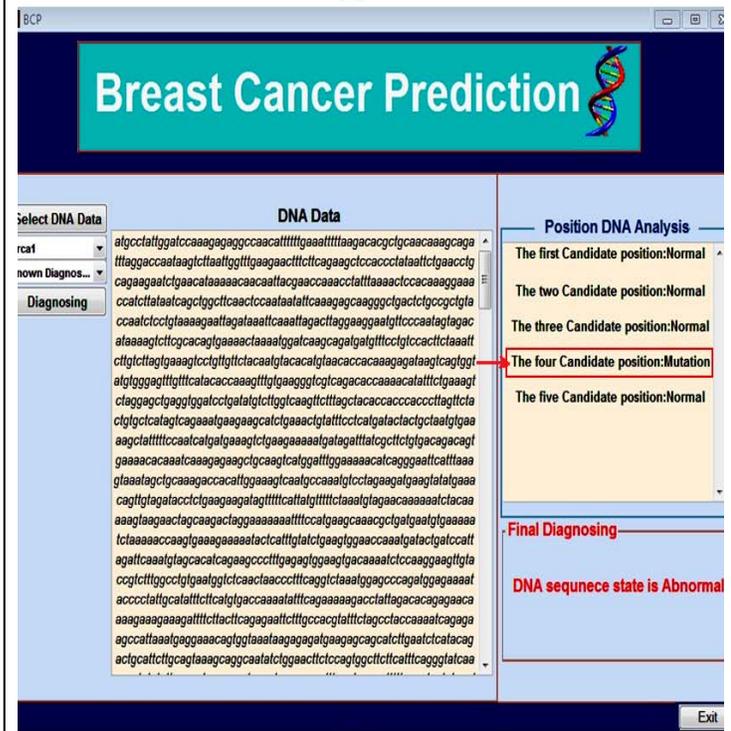

Figure 9: Shows (A) the result of test BP algorithm when there aren't mutations, while (B) when there is certain mutation

V. CONCLUSIONS AND FUTURE WORK

The main conclusions which obtained from implementing the proposed novel method of disease prediction are:

1.  The proposed novel method is first prediction method gives strong results, because it not only based on malignant mutations of genes which caused the disease, but also base on their proteins.

2.  This novel method suggested a general prediction method based on mutational in genes caused the disease, i.e. can implement this novel method for any disease when the mutations of its gene which caused the disease are known.

3.  Offers an automatic, cost effective and friendly diagnosis system for detecting malignant mutation of breast cancer as shown in Table 1, i.e. can use by any researcher or patient who needed to test malignant mutations at genes which caused breast cancer.

4.  The model of classification malignant mutations for breast cancer was developed successfully using feed-forward back-propagation neural network, i.e. obtained a best classifier for prediction a breast cancer or any other disease with mean square error 0.0000001.

As future work need to establishing a local database (e.g. for middle east) containing information about the history of each family with respect to genetic diseases and building an integrated system to enable early detection of genetic disease based on this novel method for prediction, to help as much as possible, save lives and offers advice to people in due course.

## AUTHORS PROFILE

**Ayad Ghany Ismaeel** received MSC in computer science from the National Center of Computers NCC- Institute of Postgraduate Studies, Baghdad-Iraq, and Ph.D. Computer science in qualification of computer and IP network from University of Technology, Baghdad- Iraq. 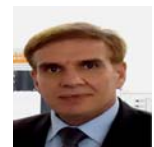

He is professor assistant, at 2003 and currently in department of Information System Engineering in Erbil Technical College- Hawler Polytechnic University (previous FTE Erbil), Iraq. His research interest in mobile, IP networks, Web application, GPS, GIS techniques, distributed systems, bioinformatics and bio-computing. He is senior lecturer in postgraduate of few universities in MSC and Ph.D. courses in computer science and software engineering as well as supervisor of multiple M.Sc. student additional Higher Diploma in Software Engineering and computer from 2007 till now at Kurdistan-Region, IRAQ.

Ayad Ghany Ismaeel is Editorial Board Member at International Journal of Distributed and Parallel Systems IJDPS http://airccse.org/journal/ijdps/editorial.html, Program Committee Member of conferences related to AIRCC worldwide, and reviewer in IJCNC (which is listed as per the Australian ARC journal ranking http://www.arc.gov.au/era/era_2012/era_journal_list.htm), other journals at AIRCC like IJDPS, IJCSIT etc (http://airccse.org/journal.html), and conferences within AIRCC http://airccse.org/, as well as he adviser and reviewer in multiple national journals. The published papers were in international and national journals about (20) paper.

**Anar Auda Ablahad Kasyounan** is graduated from software engineering of Mosul University, and now she is MSc student at research stage in computer science at Zakho University.

She is work at department of electrical and computer, school of engineering, faculty of engineering and applied science, university of Dohuk, Iraq.

Her researches interest in the fields of bioinformatics and programming software.